# Supply Chain Insecurity: Exposing Vulnerabilities in iOS Dependency Management Systems

David Schmidt[1,2], Gabriel K. Gegenhuber[3], Edgar Weippl[1], and Sebastian Schrittwieser[1,2]

[1] University of Vienna, Faculty of Computer Science
[2] Christian Doppler Lab for Assurance and Transparency in Software Protection
[3] Interdisciplinary Transformation University (IT:U)

**Abstract** Dependency management systems are a critical component in software development, enabling the incorporation of existing functionality. However, misconfigurations and malicious actors in these systems pose security risks, leading to supply chain attacks. Despite the widespread use of smartphone apps, the security of dependency management systems in the iOS software supply chain has received limited attention. In this paper, we focus on CocoaPods, one of the most widely used dependency management systems for iOS app development, but also examine the security of Carthage and Swift Package Manager (SwiftPM). We demonstrate that iOS apps expose internal dependency names and versions. Attackers can exploit this information to register previously unclaimed dependencies in CocoaPods, enabling remote code execution (RCE) on developer machines and build servers. Additionally, we show that attackers can compromise dependencies by reclaiming abandoned domains and GitHub URLs. Analyzing 9,491 real-world iOS applications, we find that a significant fraction is vulnerable to these attacks, putting millions of users at risk, including applications from major organizations for which we confirmed exploitation via responsible disclosure. Finally, we compare iOS dependency management systems with Cargo, Go modules, npm, pip, and Maven, and discuss conceptual issues and practical mitigation strategies.

## 1 Introduction

Libraries and dependency management systems form the backbone of software development by allowing developers to integrate existing functionality efficiently. Public dependency repositories host millions of libraries [70]. Due to transitive dependencies, installing a single library can implicitly introduce dozens of additional libraries [75]. Each included dependency expands the attack surface, as demonstrated by incidents such as the event-stream compromise [21] and the XZ Utils backdoor [1]. Therefore, a single malicious library can compromise all dependent projects and any device running software that integrates the malicious dependencies. These attacks exploit developers' trust in the software supply chain and are therefore difficult to detect. For instance, in the SolarWinds compromise, attackers infiltrated the update server and distributed malicious code to over 18,000 customers, which was undetected for at least nine months [69].

Researchers have investigated software supply chain risks through dependency management systems in the past. Zimmermann et al. [75] analyzed the npm ecosystem and demonstrated the cascading impact of vulnerable and malicious libraries through direct and transitive dependencies. Zahan et al. [72] identified weaknesses in npm, such as the use of abandoned domains for maintainer email addresses, which can facilitate supply chain attacks. Ohm et al. [53] discovered 174 malicious libraries distributed across npm, PyPI, and RubyGems. Birsan [4] showed that dependency names can leak online, enabling dependency confusion attacks that result in Remote Code Execution (RCE). Furthermore, Gu et al. [30] identified twelve threats to dependencies hosted on five dependency repositories and their mirror servers.

However, prior work has not analyzed software supply chain risks in the context of iOS apps, despite the deep integration of mobile apps into daily life [61] and the dominant iOS market share of 58.68% in the United States [63]. We focus our analysis on CocoaPods, a widely used dependency management system in the iOS ecosystem. Although CocoaPods announced a transition to read-only mode (starting in December 2026) as part of its gradual phase-out [67], its relevance remains significant. The project reports over 100,000 libraries and usage in more than 3 million iOS and macOS applications. Given this large installed base, thousands of apps are expected to continue relying on CocoaPods-managed dependencies in the coming years. As a result, vulnerabilities in its dependency resolution and distribution mechanisms remain highly impactful.

Mobile apps are distributed to end-user devices. Consequently, attackers can reverse engineer them, revealing insights into an app's supply chain, a scenario known as Man-at-the-End (MATE) [15]. We leverage this property to show that iOS apps expose dependency information, enabling dependency confusion attacks that can result in two primary attack vectors. First, a malicious dependency can be introduced into an app, allowing attackers to distribute malicious code to end-user devices through otherwise trusted software. Second, installing such a dependency can trigger code execution, leading to RCE. This build-time code execution enables attackers to either modify the application's source code beyond the injected dependency or to compromise the build server or developer workstation.

To the best of our knowledge, we are the first to study supply chain attacks on iOS apps. To do so, we evaluate attack surfaces in dependency management systems, focusing on technical issues rather than social engineering techniques, and answer *RQ1: Which supply chain issues do iOS dependency management systems face?* After gaining an understanding of supply chain attacks, ranging from dependency confusion to dependency hijacking attacks, we measure the impact of the discovered vulnerabilities on a dataset of 9,212 iOS apps and open-source projects on GitHub to answer *RQ2: How many iOS apps are vulnerable to supply chain attacks?* Finally, to better understand existing defense mechanisms against the discovered vulnerabilities, we evaluate the security properties of five additional dependency management systems and answer *RQ3: How can dependency management systems protect against the identified vulnerabilities?*

In this paper, we show that iOS apps reveal internal library names, which can lead to RCE on developer machines and build servers. We demonstrate this attack and responsibly disclosed vulnerabilities in mobile apps from nine companies, including Microsoft. In addition to analyzing internal (unclaimed) dependencies, we highlight the risks posed by abandoned domains and GitHub namespaces. Attackers can exploit these to hijack public iOS dependencies used by 162 apps in our dataset, affecting millions of users globally. By expanding our study to five additional dependency management systems: Cargo, Go modules, npm, pip, and Maven, we discuss mitigation strategies and highlight that the identified issues are not confined to the iOS ecosystem. In summary, we make the following key contributions:

- We show that iOS apps often leak internal dependency names and versions, which can lead to RCE on developer machines and build servers when dependency management files are misconfigured. We demonstrated the Proof of Concept (PoC) attack against apps from nine companies;
- We demonstrate that dependencies are vulnerable to hijacking attacks, affecting 162 apps (1.76%) in our dataset;
- We analyze five additional dependency management systems to contextualize our findings and evaluate how they address the identified vulnerabilities. Our findings further show that Go and npm share similar conceptual weaknesses;

For reproducibility and future work, we publish our code to find vulnerable dependencies at: `https://github.com/CDL-AsTra/supply-chain-insecurity`.

## 2 Foundations and Threat Model

In this section, we describe the recurring properties of dependency management systems and provide an overview of attack scenarios analyzed in this paper.

### 2.1 Dependency Management Systems

Dependency management systems have two aspects that are crucial for the attacks we study in this paper: (1) the location where dependencies are hosted and (2) the mechanisms used to manage library ownership.

**Dependency Location** Dependency management systems follow two models for hosting libraries: centralized or decentralized. In the centralized model, dependencies reside in a central repository. For instance, CocoaPods [11] provides a GitHub repository that enables developers to discover and integrate libraries by specifying their names. In contrast, SwiftPM [65] implements a decentralized model that requires developers to specify the location of each dependency explicitly. These locations can be provided as Uniform Resource Locators (URLs) to Git repositories, e.g., hosted on GitHub, or as local file paths.

Since not every dependency should be publicly accessible, dependency management systems with a central repository typically provide mechanisms to host private dependency repositories or specify dependency locations.

**Dependency Ownership** The second key aspect is the ownership model of the dependencies. In centralized systems, the management system is responsible for the ownership, making security dependent on the correctness and robustness of its implementation. In decentralized systems, ownership is determined by control over the URL that hosts the dependency. When dependencies are hosted on GitHub, for example, ownership depends on GitHub's URL resolution.

### 2.2 Attack Scenarios

In this section, we present supply chain attacks targeting dependency management systems.

**Dependency Confusion** Dependency confusion attacks were introduced by Birsan [4], who demonstrated vulnerabilities in pip [58] and npm [52]. In this attack, an adversary registers a dependency in a central repository using the same name as a private dependency.

If the dependency manager prioritizes public sources over private ones, or is misconfigured, it resolves the dependency to the attacker-controlled version. This can lead to RCE on developer machines or build systems that install the dependency, provided the system permits code execution during installation.

**Dependency Hijacking** Another threat is dependency hijacking, where attackers gain ownership of existing (public) libraries. Its feasibility depends on the library's hosting location. In centralized repositories, the security depends on the repository's ownership implementation. In decentralized settings, the critical factor is control over the URLs that host the dependencies.

*Abandoned GitHub URLs* An attack vector for hijacking dependencies stems from URL ownership [30]. As many dependencies are hosted on GitHub, its URLs management plays a critical role.

GitHub allows users to rename or delete their accounts. When a user renames an account, GitHub redirects the old repository URL to the new one. For example, if a user called `RAID_2025` owns the repository `proc` and renames the account to `RAID_2026`, then `https://github.com/RAID_2025/proc` automatically redirects to `https://github.com/RAID_2026/proc`. GitHub refers to the username as a namespace and the repository name as the image-name. We use this terminology throughout the paper for consistency. This redirect remains active until another user registers the previously abandoned namespace `RAID_2025` and creates a repository with the same image-name `proc`. At that point, the URL `https://github.com/RAID_2025/proc` resolves to the new repository.

If a dependency management system uses this URL to fetch the library, the new user can inject malicious code into the library. From a usability perspective, the redirect mechanism prevents broken links and supports seamless transitions. However, from a security standpoint, it obscures dependency hijacking attacks by preserving apparent functionality even after the namespace has changed.

To mitigate this risk, GitHub recommends updating all URLs after renaming a namespace. Additionally, in 2018, GitHub announced it would retire namespaces linked to repositories that had received at least 100 clones within a week [25]. In 2023, this policy was extended to retire any namespace and image-name combination whose associated image has more than 5,000 downloads [24].

*Abandoned Domains* Dependencies can be hosted on custom domains. If such a domain expires and becomes available (i.e., an abandoned domain), an attacker can register it and take control of the library. Unlike GitHub URL hijacking, where redirects preserve repository availability, libraries hosted on custom domains become temporarily unavailable during the period between expiration and hijacking, increasing the likelihood of detection.

**Typosquatting** Typosquatting attacks exploit typographical errors when developers specify dependency names [50,66,75]. Attackers register similarly named dependencies to trick developers into installing malicious ones (e.g., grammatical or semantic modifications [50]). Unlike the previous attacks, typosquatting relies on human error and is closer to social engineering, as seen in incidents such as the XZ Utils backdoor [1] and the event-stream compromise [21]. We consider typosquatting out of scope, as it does not exploit technical vulnerabilities.

### 2.3 Threat Model

In our threat model, we assume attackers have access to apps distributed through the Apple App Store and publicly available library information, such as metadata from public dependency repositories or GitHub.

Attackers target either (i) build systems that incorporate malicious dependencies during the build process or (ii) end-user devices that execute apps containing compromised libraries. By compromising a build system, attackers aim to achieve RCE in trusted development or CI environments, move laterally within the development infrastructure, or obtain sensitive assets, such as API keys or signing credentials. By compromising end-user devices, attackers aim to distribute and execute malicious functionality at scale through standard distribution channels.

Attackers exploit technical vulnerabilities without relying on social engineering. We exclude social engineering because its success primarily depends on human behavior, which we consider outside the scope of this paper.

```
1 source 'https://cdn.cocoapods.org/'
2 source 'https://github.com/Artsy/Spec'
3 use_frameworks!
4
5 target :MyApp do
6 project 'MyApp.xcodeproj'
7     pod 'GoogleAnalytics'
8     pod 'OCMock', '~>2.27'
9     pod 'Aerodramus', '2.1.4'
10    pod 'Artsy+UIFonts', '2.1.4'
11 end
```

Figure 1: A `Podfile`, analogous to `package.json` in npm, that enables dependency confusion by mixing public and private dependencies. We adapt the example from the CocoaPods `Podfile`, which we identified as vulnerable [10].

## 3 iOS Dependency Management and Threat Vectors

In this section, we analyze dependency management systems used by iOS apps to answer *Which supply chain issues do iOS dependency management systems face?*. We study CocoaPods, Carthage, and SwiftPM. CocoaPods relies on a centralized repository and is used in over 3 million apps [11], whereas Carthage [7] and SwiftPM [65] follow a decentralized design.

### 3.1 CocoaPods

**Dependency Confusion Attacks** CocoaPods supports both internal and external repositories, making it vulnerable to dependency confusion when public and private sources are mixed without explicit scoping.

Figure 1 shows a vulnerable `Podfile`, based on the official CocoaPods documentation [10]. Line one (`L1`) specifies the public CocoaPods repository, and line two adds an internal repository. Their order is critical as it determines dependency resolution precedence. `L7-L8` declare public dependencies, followed by private dependencies retrieved from the internal repository. If an attacker registers private dependencies in a public repository with matching versions, developers updating dependencies may install attacker-controlled code. `Podfiles` support wildcard versions. In Figure 1, `GoogleAnalytics` specifies no version and thus resolves to the highest available one. The wildcard for `OCMock`, '~> 2.27', leaves the patch version open, which selects the highest matching patch release.

After the initial build, CocoaPods records resolved dependencies in a `Podfile.lock` file, which is reused in subsequent builds. Dependency confusion can occur when dependencies are explicitly updated (i.e., using the `update` command rather than `install`, which relies on cached dependencies). Developers can address the vulnerability in two ways. First, by reordering the sources, as the order controls CocoaPods resolution priority. Second, by explicitly specifying the internal dependency location, as done in `L10`. Since our example builds on the `Podfile` from the official CocoaPods documentation [10], we reported their example as vulnerable and recommended updating it accordingly.

Dependency confusion attacks enable attackers to gain RCE on developer machines and build servers via the `prepare_command`, which allows the execution of arbitrary shell code. To directly compromise the mobile app's code through dependency confusion, an attacker must publish a functional malicious dependency that implements the expected interfaces to avoid detection. Dependency confusion arises when private dependencies are used, whose code is typically unavailable, requiring attackers to obtain it (e.g., via RCE or reverse engineering). As this is difficult, attacks typically target build systems.

*Finding Internal Dependencies* Mobile apps are distributed to end-users, placing them in a MATE setting where attackers can extract dependency names and versions from published apps. While prior work identified such information in npm or pip by analyzing leaked or embedded dependency files [4], iOS apps directly expose it through their directory structure and configuration files.

CocoaPods supports two integration modes: *dynamic frameworks* and *static libraries*. When a `Podfile` includes the `use_frameworks!` directive, CocoaPods builds each dependency as a separate dynamic framework. These frameworks appear in the app bundle under the `Frameworks` directory, following the naming scheme `{framework_name}.framework` [16].[1] Within each `.framework` directory, CocoaPods includes an `Info.plist` which reveals detailed framework information. By default, `CFBundleVersion` exposes the dependency's version and `CFBundleIdentifier` prefixes the library name with `org.cocoapods`, thereby also revealing the use of CocoaPods.

When dependencies are integrated as static libraries by omitting `use_frameworks!` they are linked directly into the main binary and no separate `.framework` bundle appears in the app package. Version and naming information may still be recoverable through binary fingerprinting techniques, though this is less straightforward and accurate than in the dynamic framework case. We focus primarily on the dynamic framework setting in the remainder of this section, as it provides the most direct and reliable source of dependency metadata.

**Dependency Hijacking** CocoaPods implements its own user management, which requires authentication and authorization to create and update dependencies. CocoaPods does not use password-based authentication; instead, it sends a magic login link to the registered email address. Consequently, accounts associated with email addresses from abandoned domains are vulnerable to hijacking. This threat is further amplified as new versions can be published silently, even when multiple accounts own a dependency. Typically, all dependency owners receive a notification email when a new library version is published or an old version is deleted. However, neither removing nor adding an account as a library owner triggers notifications. Thus, attackers could first remove other owners from a library, second, introduce malicious changes, and later re-add the original owners unnoticed.

[1] Although this convention can be overridden via the `module_name` or `header_dir` fields, it is not common in practice (as we show in Section 4).

CocoaPods does not host libraries directly. Instead, it fetches them from external URLs, allowing attackers to hijack dependencies via compromised sources (e.g., abandoned domains or GitHub namespaces). Once a dependency is installed, the `Podfile.lock` stores the library version and the checksum of its specification. However, since the checksum is calculated over the specification file [12], and not the entire library, attackers can nevertheless hijack the URL and silently inject malicious code. CocoaPods caches libraries locally, thus malicious modifications are fetched from the URL only if a developer downloads the dependency for the first time or explicitly clears the cache.

CocoaPods supports specifying commit or archive hashes to mitigate hijacking [9]. While this hinders hijacking via URLs, attackers can still compromise existing library versions by taking over the library owner's account (e.g., via an abandoned email domain). With such access, an attacker can modify the specification file of an existing version, e.g., by removing the checksum and redirecting the source URL. This modification changes the specification and, consequently, the checksum recorded in `Podfile.lock`, which developers could notice. Alternatively, attackers can publish new dependency versions. Since developers fetch these versions during explicit dependency updates, a checksum change appears expected and therefore is less suspicious. Central repositories serve as indexes, providing developers with an overview of libraries. We leverage this in Section 4 to quantify the libraries vulnerable to hijacking attacks.

### 3.2 Carthage and SwiftPM

Carthage and SwiftPM do not use a central repository and require explicit dependency references, preventing dependency confusion but shifting trust to external hosting. Thus, dependencies can be hijacked via abandoned domains or GitHub namespaces. Since libraries published to CocoaPods normally reside in a Git repository, they can also be consumed via Carthage or SwiftPM. Therefore, by analyzing whether dependencies published in the central CocoaPods repository can be hijacked via an abandoned URL, we implicitly gain insights into the exposure of Carthage and SwiftPM users as well (see Section 4).

In decentralized systems, control over the URL suffices to publish updates. Although Carthage and SwiftPM record dependency versions in lock files, they provide limited protection against hijacking, as updates may still resolve to attacker-controlled code. Carthage records versions without checksums unless a commit hash is explicitly specified. SwiftPM records commit hashes, but these reflect the current state of version tags: if an attacker moves a tag to a malicious commit, the next build updates the lock file. True immutability would require developers to pin dependencies to a commit hash rather than a version tag, which Apple explicitly discourages for SwiftPM [3]. Consequently, both package managers remain susceptible to hijacking attacks on abandoned repositories.

*Takeaways*

- Information about internal dependency names and versions included in iOS apps can enable dependency confusion attacks when dependency management files are misconfigured;
- CocoaPods authentication is insecure. Library hijacking possible when owners registered with email addresses from abandoned domains;
- Abandoned GitHub namespaces can lead to hijacking attacks.

## 4 Measurement of Vulnerable iOS Apps

After discussing supply chain attacks on iOS apps, we measure the vulnerable apps and answer *How many iOS apps are vulnerable to supply chain attacks?*

### 4.1 Datasets

To obtain an overview of existing iOS libraries, we use the publicly available CocoaPods `Spec repository` [13], hosted on GitHub. The repository provides metadata for each library, including the hosting location (e.g., GitHub or an external URL) and naming and version information.

To analyze the usage of vulnerable libraries, we use a dataset of 9,212 iOS apps downloaded in 2024, introduced by Schmidt et al. [61]. We refer to this dataset as `2024`. The dataset includes matching Android versions, whose installation counts we use as a proxy for iOS popularity, since the App Store does not provide comparable metrics [61,64].

We additionally collected 279 iOS apps participating in responsible disclosure programs to demonstrate the feasibility of dependency confusion attacks. These apps belong to 105 companies listed on platforms such as HackerOne [32], Bugcrowd [5], and Intigriti [36], and to well-known companies such as Microsoft [48] and Google [27], each with its own responsible disclosure program. We refer to this dataset as `disclosure 2025`.

Further, we query the GitHub API for repositories that build upon vulnerable libraries. To this end, we search public repositories for Carthage, CocoaPods, and SwiftPM dependency files, which yields approximately[2] 57,000 repositories.

### 4.2 Dependency Confusion in CocoaPods

**Methodology** To identify apps vulnerable to dependency confusion attacks, we implemented a Python-based analysis that extracts all framework names from iOS apps and retrieves their versions and identifiers from the `Info.plist` files. We then compare the extracted information with the publicly available CocoaPods from the `Spec repository` [13]. The specification repository also reveals naming practices of CocoaPod libraries, such as if developers overwrite default framework directory names using `header_dir` or `module_name`. Thus, it indicates whether our approach to extract dependency names from the framework directory is effective.

[2] We separately search for `Cartfile` (6,928), `Package.swift` (9,924), and `Podfile` (40,328) without deduplicating results due to API limitations of GitHub.

To demonstrate feasibility and enable responsible disclosure, we targeted apps that participate in responsible disclosure programs (i.e., the `disclosure 2025` dataset). To mimic a bad actor, we automatically published libraries with our university's email addresses. These libraries included a `README` describing the research purpose, contact information, and stated our limited deployment period for the study.

When our PoC was installed, we collected the hostname, installation directory, dependency name, and external IP address via Domain Name System (DNS) lookups to confirm execution within company environments.[3] Successful execution demonstrates that an attacker could achieve RCE on the developer machine or build server that installs the dependency. The collected information also enabled us to automatically remove the test libraries to avoid disruption of the build processes of the affected company.

**Results** To examine framework naming practices, we analyzed whether libraries specified a `header_dir` or `module_name` in at least one version. Both fields overwrite the default framework metadata, which otherwise matches the library name. Out of 99,261 libraries that were identified across the `Spec repository` dataset, we found that 95,741 libraries (96.45%) did not specify either, 3,155 libraries (3.18%) specify only `module_name`, 234 libraries (0.24%) only `header_dir`, and 131 libraries (0.13%) both. Only 1,689 libraries (1.70%) overwrite the default name with a name that differs from the library name. Since all remaining libraries follow the default naming convention, this supports the accuracy of our metric for deriving library names from framework metadata.

To obtain insights into the use of static and dynamic linking with CocoaPods, we further search for the dynamic linking keyword `use_frameworks!` in public GitHub projects, identifying it in 46%.[4] We assume that statically linked apps also contain vulnerable libraries. Consequently, the number of affected libraries exceeds those detected by our analysis. However, our focus on dynamically linked libraries helps avoid False Positive (FP), as they provide reliable metadata.

*CocoaPod Usage* We identified at least one framework directory in 7,981 apps (86.64%) of the `2024` dataset, of which 5,097 apps (55.33%) use identifiers prefixed by `org.cocoapods`. Comparing framework names against the public CocoaPods repository revealed 9,866 unregistered frameworks across 2,084 apps (22.62%), indicating potential vulnerability to dependency confusion. However, due to ethical considerations, we did not publicly register those frameworks and refrained from any actual exploitation attempts. Note that, without executing the attack, we cannot determine the order in which the build system resolves dependency sources. Therefore, the 22.62% of apps represent an upper bound on the number of vulnerable apps with dynamically linked libraries.

[3] We used DNS lookups to reduce the likelihood that firewalls blocked the communication [4,57].

[4] The search for `Podfile` returns 40,328 projects, while adding the keyword filter yields 18,688 results.

Table 1: Overview of analyzed *libraries* and *apps*. *Candidates* denotes the total number of relevant entities for each attack scenario (e.g., library owners, or domains), while *targets* captures the number of vulnerable CocoaPods. For vulnerable apps, we use the Android installation count as a proxy for app popularity, reported as *MAD*.

| | **Libraries** | | | **Apps** | |
|---|---|---|---|---|---|
| | **Candidates** | **Vulnerable** | **Targets** | **# Apps** | **MAD** |
| Dep. Confusion | 24,838 | 9,866 (39.7%) | | 2,084 (22.6%)[a] | 499,900 |
| Owner Hij. | 8,243 | 107 ( 1.3%) | 213 | 97 ( 1.1%) | 499,990 |
| GitHub Hij. | 83,632 | 5,056 ( 6.1%) | 5,137 | 80 ( 0.9%) | 499,945 |
| Source Hij. | 1,969 | 83 ( 4.2%) | 127 | | |

[a]The proportion of apps vulnerable to dependency confusion represents an upper bound because we lack information about the dependency source order. We demonstrated the attack on apps from 8.57% of evaluated companies with disclosure programs.

*Responsible Disclosure-based Exploitation* Among apps with explicit disclosure programs (`disclosure 2025` dataset), 116 (41.58%) contained frameworks with CocoaPods identifiers, corresponding to 60 companies (57.14%). A total of 764 frameworks were not registered in the central repository, originating from 67 apps (24.01%) across 34 companies (32.38%).[5]

Overall, our PoC succeeded against internal libraries from nine companies (8.57%), including Microsoft. Thus, we received DNS lookups from company-related environments, demonstrating that an attacker could achieve RCE in these environments. Additional apps in our test set may also be vulnerable. However, dependency confusion requires an affected project to execute the CocoaPods `update` command, which likely did not occur during our testing period. This requirement may explain why our attack primarily succeeded against apps from larger companies, which presumably update their dependencies more frequently.

*Exploitation in the Wild* After we launched our PoC at the end of February 2025, we observed another party deploying CocoaPods targeting dependency confusion by the end of March 2025 [17]. We contacted the user via the email address associated with the published library to clarify their intentions and identity. In response, the user stated that their motivation was to earn bug bounties and referenced a video on dependency confusion attacks [51] which explained the attack demonstrated by Birsan [4]. We are unaware of any other malicious or non-malicious exploitation in the CocoaPods ecosystem.

[5] We did not test all potential vulnerable libraries of an app due to ethical considerations. Instead, we deployed a subset of libraries per company at the same time and stopped testing the company's dependencies once we identified a vulnerability. Thus, we reported results at the company level instead of the app level.

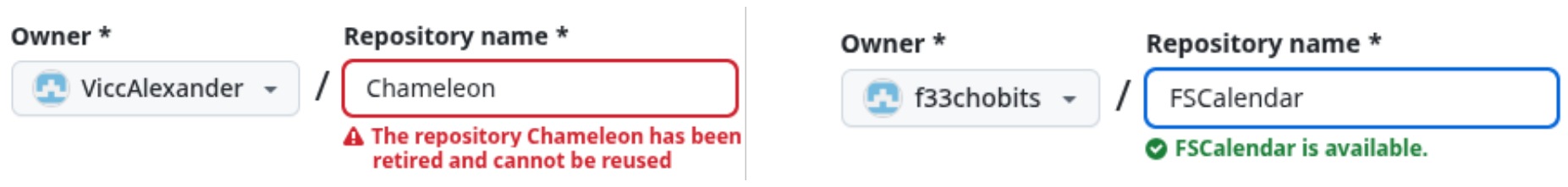

Figure 2: GitHub displays if the repository is available or retired before creating it. Since we do not create the repository, the URL forwarding remains functional.

Positively, this contributed to CocoaPods developers' decision to reject new libraries that contain `prepare_commands` [67] in May 2025, thereby mitigating the threat of RCE during dependency installation. However, attackers can still infect user devices if they have access to the dependency code or can successfully reverse-engineer it to provide a functional, but malicious, dependency.

### 4.3 Dependency Hijacking

**Methodology** To assess dependency hijacking risks from abandoned domains and GitHub namespaces, we extracted (1) the list of available libraries and (2) their associated dependency hosting locations. Additionally, we collected their integrity verification checks from the CocoaPods specification repository. Unlike dependency confusion, which requires extracting internal dependency names from app bundles, dependency hijacking targets public dependencies and can therefore be analyzed directly via the registry.

We obtained the email addresses of all library owners to analyze the potential for hijacking CocoaPod libraries through abandoned email addresses. Since this information is not available in the specification repository, we queried the Application Programming Interface (API) at `https://trunk.cocoapods.org/api/v1/pods/{name}` to obtain the library owners and their email addresses. We subsequently verified the availability of email domains using the bulk domain search services provided by GoDaddy [26] and Namecheap [49], cross-validating the results.

To identify abandoned hosting locations, we differentiated between dependencies hosted on GitHub and those hosted on custom domains. Similar to the approach for abandoned email domains, we again checked the availability of each hosting domain. For libraries hosted on GitHub, we initially queried the GitHub API to verify the existence of each namespace. If a repository redirected to a new URL, we also recorded the number of stars for the redirected repository.

Afterward, we manually registered available GitHub namespaces for abandoned repositories and entered the corresponding image names during repository creation. Since we did not finish creating the repository but only checked its availability this step did not break the forwarding, see Figure 2. However, it allowed us to evaluate whether GitHub has retired the URL. We further extracted commit and archive hashes listed in the specification, both of which can prevent malicious modifications of a library if attackers hijack the associated URLs. Finally, we applied our analysis described in Section 4.2 to identify the usage of vulnerable libraries in iOS apps. In addition to evaluating the `2024` dataset, we analyzed iOS projects published on GitHub by identifying vulnerable dependency usages in CocoaPods, Carthage, and SwiftPM projects.

We deliberately refrained from applying the full exploit chain to real libraries, as registering an abandoned namespace irreversibly breaks existing redirects, potentially disrupting the build processes of all apps that depend on them. We provide more details on ethical considerations in Section 6. However, to demonstrate the real-world feasibility of the described hijacking attack, we mimic the full exploit chain using controlled experiments. For example, to demonstrate the breaking of GitHub's redirect mechanism, we created a fresh namespace, renamed it to establish a redirect, and then registered the original namespace to confirm that this permanently disables redirects and that the dependency management system is vulnerable.

### Results

*Abandoned Email Domains* Of 99,261 CocoaPods referenced in the `Spec repository`, we extracted 8,243 domains of owner email addresses, of which 107 were abandoned (1.30%) owning 213 libraries, see Table 1. In our dataset, 97 apps (1.05%) use at least one vulnerable dependency.

The most common vulnerable library was `DZNEmptyDataSet` (found in 63 apps, 0.68%), followed by `SlackTextViewController` (15 apps, 0.16%) and `DTTJailbreakDetection` (8 apps, 0.09%).

In addition to analyzing the apps from the `2024` dataset, we evaluated 40,328 public repositories that we retrieved via GitHub search. We found 121 GitHub repositories relying on vulnerable libraries, as detailed in Table 2. The library `SlackTextViewController` was the most prevalent (appearing in 15 repositories), followed by `UIColor-HexString` (14 repositories), and `DTTJailbreakDetection` (13 repositories). Notably, `DZNEmptyDataSet`, the most common vulnerable library in apps, ranked ninth on GitHub with six occurrences. Also, popular open-source repositories used vulnerable dependencies, with an average star count of 169, and one repository even had 7,129 stars.

To mitigate the imminent threat of available domains, we proactively registered two abandoned email domains associated with the three most frequently encountered dependencies in iOS apps. Since `DZNEmptyDataSet` and `DTTJailbreakDetection` share the same owner. We did not alter library specifications or ownership and registered them only to prevent malicious hijacks.

*Abandoned GitHub Namespaces* We analyzed the GitHub URLs referenced by CocoaPod libraries. Of 83,632 unique GitHub repositories, 5,056 (6.05%) were abandoned. We manually verified the possibility of registering the namespace and image-name for every dependency that forwarded to an image with at least 100 stars (236), and for all remaining URLs used by libraries discovered in an app in our dataset (31). Of repositories with at least 100 stars, we would have been able to register 186 (78.81%). Alarmingly, among vulnerable GitHub URLs, 42 repositories have over 1,000 stars. We would have assumed that these repositories had reached the 5,000-download threshold required for retirement.[6]

Table 2: Number of GitHub repositories using vulnerable dependencies. The column *vulnerable* indicates where the vulnerability resides, i.e., in the *owner* domain, the *GitHub* URL, or the *source* domain.

| System | Vulnerable | # Repos. | Stars | | |
|---|---|---|---|---|---|
| | | | **Avg.** | **Std.** | **Max.** |
| CocoaPods | Owner | 121 | 169.1 | 867.4 | 7,129 |
| CocoaPods | GitHub | 671 | 25.7 | 222.3 | 3,805 |
| Carthage | GitHub | 11 | 2.8 | 2.6 | 7 |
| SwiftPM | GitHub | 43 | 43.3 | 210.5 | 1,400 |
| CocoaPods | Source | 4 | 1.3 | 2.2 | 5 |
| Total | | 834 | 42.3 | 370.7 | 7,129 |

We identified 5,137 vulnerable CocoaPods libraries, exceeding the number of abandoned GitHub namespaces, as 93 were referenced by multiple libraries. We did not observe any malicious intent behind this behavior. Instead, library authors renamed the library or used different names to reference specific branches and builds, e.g., production and debug. Among vulnerable libraries, 49 (0.95%) specified a commit hash in at least one version, while 33 (0.64%) did so across all versions, indicating that such countermeasures are rarely used in practice.

In our `2024` dataset, 80 apps (0.87%) relied on vulnerable CocoaPod libraries, including three apps with over 100 M installations: two fitness apps, and one antivirus app. In total, we identified the use of 49 distinct vulnerable libraries. With seven occurrences, `JTMaterialSpinner` is the most popular, followed by `JXPageControl` with six uses. Remarkably, we did not find any app using an abandoned GitHub URL that protected the library by specifying its hash.

We detected insecure CocoaPods dependencies in 671 GitHub repositories. In two additional cases, the library referenced an abandoned GitHub URL, but the library version specification included a commit hash, making the projects not vulnerable. Additionally, we identified references to abandoned GitHub URLs in 11 Carthage repositories and 43 SwiftPM repositories, as summarized in Table 2.

*Abandoned Source Locations* We extracted 1,969 domains from CocoaPods specifications, of which 83 (4.22%) were abandoned. 127 libraries referenced these abandoned domains. Notably, libraries occasionally changed hosting locations across versions (e.g., from GitHub to project domains). 25 vulnerable libraries referenced multiple hosting locations: 13 libraries used vulnerable domains in earlier versions, while 7 transitioned from secure to abandoned domains in their recent versions. Only one of these 25 libraries applied an integrity check. Nevertheless, it does not mitigate the issue, as none of the versions that can be hijacked specified a check. Three libraries contained vulnerable versions, but none were used in affected apps. On GitHub, we found four repositories that referenced vulnerable CocoaPods versions, while no vulnerable domain references appeared in SwiftPM or Carthage configuration files.

[6] Possibly they were renamed before reaching GitHub's required download threshold.

Table 3: Dependency confusion characteristics across systems.

| **Framework** | **Private Dependencies** | **Code Exec.** | **Dep. Confusion** |
|---|---|---|---|
| CocoaPods | Implicit | Yes | ✓ |
| Carthage | Explicit | No | 7 |
| SwiftPM | Explicit | No | 7 |
| Cargo | Explicit | No | 7 |
| Go modules | Explicit | No | 7 |
| npm | Implicit | Yes[a] | ✓ |
| pip | Implicit | Yes | ✓ |
| Maven | Explicit | No | 7 |

[a]Starting with version 12, released in July 2026, npm disables code execution by default [23].

***Takeaways***

- We showed that in our dataset 2,084 apps (22.62%) are potentially vulnerable to dependency confusion attacks. We successfully demonstrated the attack on apps from nine companies with disclosure programs, including Microsoft;
- We discovered 213 CocoaPod libraries vulnerable to hijacking attacks because their owners registered with abandoned email addresses;
- We identified 5,056 vulnerable abandoned GitHub URLs whose dependencies were used by 80 apps (0.87%). This highlights that GitHub's retirement policy is insufficient.

## 5 Dependency Management Comparison and Mitigations

After analyzing iOS apps vulnerable to supply chain attacks, we answer *How can dependency management systems protect against the identified vulnerabilities?* Both dependency confusion and dependency hijacking are not limited to the iOS ecosystem; they have also been observed in other widely used dependency management systems, which have introduced various countermeasures in response.

To contextualize our findings and derive effective mitigation strategies, we examine Cargo, Go modules, npm, pip, and Maven. These ecosystems are widely adopted and represent different design approaches, enabling a comparative analysis of their security mechanisms.

Table 4: Dependency hijacking attack vectors and mitigations across dependency management systems. A checkmark indicates practical feasibility under current security mechanisms; ∼ denotes conditional risk. For Maven, the three hijacking vectors largely overlap, as authentication is based on control over a domain or a GitHub account.

| **Framework** | **Registry** | **Owner Hijack** | **GitHub Hijack** | **Domain Hijack** | **Mitigation** |
|---|---|---|---|---|---|
| CocoaPods | Specs Repository | ✓ | ✓ | ✓ | Commit pinning |
| Carthage | – | – | ✓ | ✓ | Commit pinning |
| SwiftPM | – | – | ✓ | ✓ | Commit pinning |
| Cargo | crates.io | 7 | 7 | ∼ | 2FA (GitHub mandatory) |
| Go modules | Go proxy | – | ✓ | ✓ | Proxy cache; commit pin. |
| npm | npm registry | ✓ | ✓[a] | ✓[a] | 2FA (optional) |
| pip | PyPI | 7 | ✓ | ∼ | 2FA (mandatory); checks |
| Maven | Maven Central | ✓ | ✓ | ✓ | DNS verification |

[a]Starting with version 12, released in July 2026, npm disables dependency downloads from GitHub and remote URLs by default [23].

*Dependency Confusion* While Appendix A.1 provides detailed insights into each dependency management system, Table 3 summarizes their susceptibility to dependency confusion, considering both implicit resolution of private dependencies and support for code execution during dependency installation. In addition to CocoaPods, npm, and pip are vulnerable, while systems requiring explicit dependency sources are not. Moreover, not supporting or allowing code execution during dependency installation eliminates this attack vector towards build machines and constitutes an effective mitigation.

Dependency confusion requires private dependency names, e.g., by guessing or extracting them from binaries. In addition to demonstrating that such names are exposed for dynamic frameworks in CocoaPods, we show in Appendix A.2 that npm dependencies are also used within iOS applications (in the 2024 dataset in approximately 9% of apps) and similarly leaks dependency names via directories such as `node_modules/` and `www/plugins/`. This can enable confusion attacks affecting 27 apps, including one with over 50 million installations. Starting with npm version 12, released in July 2026, npm disables code execution by default, thereby reducing the risk of RCE through dependency confusion.

In summary, to prevent dependency confusion, systems should require developers to specify private dependency sources explicitly, as Cargo does. Additionally, they should disable code execution by default during dependency installation to reduce the attack surface.

*Dependency Hijacking* Table 4 summarizes the susceptibility of different dependency management systems to dependency hijacking. Appendix A.1 provides more detailed insights regarding the individual dependency management systems. Most frameworks rely on a central registry as an index of available dependencies.[7] Exceptions are Carthage and SwiftPM, which do not use a central registry but instead reference dependencies directly via their URL. Go modules conceptually follow a similar decentralized model; however, in practice, dependencies are typically retrieved via the central Go proxy, which introduces caching and exhibits characteristics of a centralized marketplace.

Dependencies can be hijacked at multiple stages. First, the dependency can be hijacked directly at the registry level (e.g., if an attacker hijacks a developer's email account/domain), which is possible for CocoaPods (magic login links), npm (email-based password reset), and Maven (via control over the corresponding `TXT` records). In contrast, crates.io and PyPI enforce two-factor authentication (2FA), which effectively mitigates registry-level hijacking.

Second, dependencies can be hijacked by taking over the corresponding GitHub namespace. This affects most frameworks, as they ultimately rely on external repository hosting. An exception is *crates.io*, which centralizes authentication via GitHub and binds packages to authenticated accounts rather than namespaces, thereby preventing hijacking through namespace reuse. For Go modules, the use of a central proxy introduces additional complexity, as cached dependencies may undermine GitHub's namespace retirement mechanism. Our analysis in Appendix A.3 shows that abandoned namespaces and domains remain exploitable, leaving even widely used projects (with more than 5,000 stars on GitHub) and those from well-known organizations (e.g., AWS and Kubernetes) vulnerable to hijacking.

Finally, dependencies can be hijacked via domain takeover when hosted on external domains. This affects all frameworks resolving dependencies via URLs. In contrast, Cargo, and PyPI are not affected in their default configurations, as dependencies are served via central repositories, but become vulnerable when dependencies are fetched from external sources such as Git repositories. To reduce this risk, npm version 12 (released in July 2026), disables dependency downloads from GitHub and remote URLs by default.

As shown by Cargo and recent changes in pip, requiring two-factor authentication (2FA) for authentication is a recommended strategy to mitigate hijacking attacks stemming from abandoned URLs. Active scanning, as performed in our work and also by pip, enables the proactive detection of abandoned domains, thereby reducing the likelihood of attacks.

Our measurements of CocoaPods and Go dependencies vulnerable to hijacking on GitHub reveal that GitHub's account retirement policy alone is insufficient. This is particularly evident with Go dependencies, where the Go proxy's caching behavior may counteract GitHub's mitigation. From a security perspective, dependency management systems should host dependencies directly rather than forwarding requests to external hosting URLs. Direct hosting allows enforcing authentication policies independently of third-party platforms. As illustrated by the Go ecosystem, this approach is only effective if the hosting platform implements a reliable authentication and authorization mechanism.

[7] Some frameworks support alternative registries (e.g., `lib.rs` instead of `crates.io`, or *JitPack* instead of *Maven Central*), introducing ambiguities that also may be exploitable for dependency hijacking.

Although we recommend directly specifying URLs for internal dependencies, doing so for public URLs reintroduces reliance on external platform security and should be avoided when possible. If unavoidable, developers should use immutable references, e.g., commit hashes, and carefully review updates. Otherwise, attackers may hijack the referenced URL to inject malicious code into newly published versions.

***Takeaways***

- Dependency management systems should require developers to explicitly specify private dependencies and disable code execution by default to prevent dependency confusion attacks;
- They should enforce 2FA and host dependencies themselves to mitigate hijacking attacks;
- And mandate commit hashes or checksums for dependencies referenced by their URLs.

## 6 Ethical Considerations

*Dependency Confusion* We demonstrated attacks only on apps with responsible disclosure programs that permitted tests through their policies. To verify that the installation originated from the targeted company, we collected the following information: (1) external IP address, (2) hostname, and (3) installation directory. This approach follows prior work on dependency confusion [4]. Further, we were asked by disclosure programs for evidence of potential compromise of the company's infrastructure, which we could provide with the collected information. We deleted all data after processing or within 31 days. We included our contact details, i.e., our university email addresses, and a brief explanation in the `.podspec`, and `README` files that accompanied each deployed library. We assessed the risk of affecting unrelated apps as minimal, since internal dependency names are typically organization-specific.

We strongly believe it is important to demonstrate attacks in practice to highlight their exploitability and potential impact. Further, it raises awareness of the vulnerability and supports fixing insecure implementations before malicious actors can exploit them. As a result, our findings contributed to the CocoaPods maintainers' decision to reject new libraries containing a `prepare_command` as of May 2025 [67], effectively mitigating RCE via dependency confusion.

*Dependency Hijacking* We did not hijack real libraries nor modified other owners' accounts of vulnerable libraries, as doing so could affect numerous apps and cause lasting damage. To demonstrate the feasibility of dependency hijacking, we validated our approach using a controlled experiment of self-owned dependencies.

To prevent attackers from registering abandoned domains and namespaces, we proactively claimed the most critical ones ourselves without interfering with their functionality. We did not register all of them due to the manual effort required for namespace creation and the cost of domain registration.

## 7 Disclosure

*Dependency Confusion* We responsibly disclosed the nine successful PoC dependency confusions. Of the reports we submitted, three companies classified them as critical, three as high, and one as medium. The remaining two companies did not provide a classification.

*Abandoned Email Domains* We disclosed our findings to both CocoaPods developers and affected apps. As of writing, we have not received any update from CocoaPods. However, in November 2024, CocoaPods announced that the central dependency repository will become read-only by the end of 2026 [67]. Once the repository enters read-only mode, the threat of dependency-hijacking attacks via abandoned owner domains will disappear. Even if an attacker successfully hijacks a dependency owner account, they will no longer be able to publish new malicious library versions to CocoaPods.

*Dependency Hijacking* We reported the attack scenario and the vulnerable repositories to GitHub. In their response, they highlighted that the forwarding mechanism is an intentional design decision. Additionally, they mentioned that they mitigate it for popular repositories using the described retirement strategy (see Section 2.2) and that they display warning messages to users when they delete or rename their namespace.

However, as our results show, this is not sufficient. To mitigate this attack vector, all previously used image-names should be retired from the namespace. Alternatively, removing the forwarding to the new namespace could help. This would cause dependencies to break after renaming, which library owners might notice (e.g., through user complaints) and could prompt updates to references.

We further reported the identified issues in CocoaPods to the maintainers and sent 154 disclosure emails to app developers using vulnerable dependencies. Following the approach by Schmidt et al. [61], we used the contact information from the Google Play Store to obtain the corresponding email addresses. We merged reports with a common email address to avoid sending multiple emails to the same developer.

*npm (cf. Appendix A.2)* We sent responsible disclosure emails using the contact addresses listed on the Google Play Store. Each message described the attack scenario, explained potential consequences, and included mitigation strategies.

*Go Modules (cf. Appendix A.3)* We responsibly disclosed the issue and the affected dependencies to Google in April 2025. Google acknowledged the report and opened two bug tickets: one concerning abandoned domains and another regarding GitHub namespaces. At the time of writing, Google closed the issue of abandoned domains as *"[...] It is unfortunate, but working as intended. [...]"*, while they left the other issue open.

## 8 Related Work

*Detection of Malicious Dependencies* Related work explored the detection of malicious libraries published to dependency repositories [18,33,34,35,54,60,73,74]. Guo et al. [31] examined the behavior of malicious pip dependencies, while Vu et al. [68] interviewed developers to understand their needs for detecting malicious Python dependencies. Taylor et al. [66] developed defenses against typosquatting prior to installation. Neupane et al. [50] explored syntactic and semantic variants related to typosquatting designed to trick developers into installing malicious dependencies. Andreoli et al. [2] analyzed known supply chain attacks, while Martínez and Durán [44] examined the SolarWinds breach.

In contrast, we focus on identifying new attack vectors within the iOS app ecosystem, rather than detecting attacks.

*Dependency Confusion and Hijacking Attacks* Zimmermann et al. [75] studied the npm ecosystem and highlighted threats posed by dependency hijacking via transitive dependencies. Birsan [4] demonstrated that leaked internal dependency names can enable RCE when dependency managers prioritize public repositories. Ladisa et al. [43] analyzed remote code execution vulnerabilities in seven dependency management systems by differentiating between install time and run time code execution. Zahan et al. [72] discovered that 2,818 maintainer domains in npm are available and that 2.2% of dependencies execute install scripts. Wyss et al. [71] proposed Latch, a system that prevents code execution during npm dependency installation. Gu et al. [30] identified twelve threat categories for dependencies hosted across six repositories and mirrors, including dependency hijacking attacks in Go. We extend this line of research to the iOS ecosystem and demonstrate that widely used apps were vulnerable to dependency confusion and hijacking attacks.

Ladisa et al. [42] summarized the risks associated with supply chain attacks and proposed mitigations, including 2FA, reviewing merge requests, and disabling install-time code execution. Other studies examined the security practices of open-source contributors [40] and investigated software signing procedures [39] to reduce attack surfaces. Sammak et al. [59] collected insights from 18 developers on the practical challenges of securing the software supply chain.

In contrast, our study directly addresses the mitigation of concrete attack vectors we identified and demonstrated in iOS dependency management systems.

## 9 Conclusion

In this paper, we demonstrated that the library names and versions revealed in iOS apps can enable dependency confusion attacks, effectively granting attackers RCE on build servers and developer devices. We further showed that attackers can exploit abandoned domains and GitHub namespaces to hijack iOS libraries.

Our measurement study of 9,212 iOS apps revealed that 162 apps (1.76%) had dependencies that could be affected by hijacking attacks, including popular apps with at least 100 million installations. Moreover, we demonstrated in practice dependency confusion attacks against apps from nine well-known companies, including Microsoft.

Additionally, we analyzed the security of five other dependency management systems: Cargo, Go modules, Maven, npm, and pip, to highlight countermeasures like 2FA and disabling code execution during installation. However, we also demonstrated that these systems face similar conceptual issues.

## Acknowledgments

The financial support by the Austrian Federal Ministry of Economy, Energy and Tourism, the National Foundation for Research, Technology and Development, the Christian Doppler Research Association is gratefully acknowledged. This research was funded in part by the Austrian Science Fund (FWF) P 33656.

## References

1. Akamai Security Intelligence Group: XZ Utils Backdoor — Everything You Need to Know, and What You Can Do, https://www.akamai.com/blog/security-research/critical-linux-backdoor-xz-utils-discovered-what-to-know, archived at https://archive.ph/NVOV0
2. Andreoli, A., Lounis, A., Debbabi, M., Hanna, A.: On the Prevalence of Software Supply Chain Attacks: Empirical Study and Investigative Framework. Forensic Science International: Digital Investigation **44** (2023). https://doi.org/10.1016/j.fsidi.2023.301508
3. Apple Documentation - Swift Package Manager, https://docs.swift.org/package-manager/PackageDescription/PackageDescription.html
4. Birsan, A.: Dependency Confusion: How I Hacked Into Apple, Microsoft and Dozens of Other Companies, https://medium.com/@alex.birsan/dependency-confusion-4a5d60fec610, archived at https://archive.ph/455ky
5. Bugcrowd, bugcrowd.com/
6. The Cargo Book, https://doc.rust-lang.org/cargo/index.html, archived at https://archive.ph/9OPy9
7. Carthage: GitHub – A simple, decentralized dependency manager for Cocoa, https://github.com/Carthage/Carthage, archived at https://archive.ph/KVMkT
8. Claburn, T.: The Register – Expert grabs expired domain for NPM package to make a point, https://www.theregister.com/2022/05/10/security_npm_email/, archived at https://archive.ph/zYi1W
9. CocoaPods: CocoaPods Guides – Podspec Syntax Reference, https://guides.cocoapods.org/syntax/podspec.html#group_subspecs, archived at: https://archive.ph/yo1tB
10. CocoaPods: CocoaPods Guides – The Podfile, https://guides.cocoapods.org/using/the-podfile.html, archived at https://archive.ph/bmXAs
11. CocoaPods: CocoaPods.org, https://cocoapods.org/, archived at https://archive.ph/yQ24y

12. CocoaPods: GitHub – CocoaPods/Core, https://github.com/CocoaPods/Core/blob/a53e235aa4d1eec8a21042e022ba7bcaca14ae56/lib/cocoapods-core/podfile.rb#L259, archived at: https://archive.ph/oDorz
13. CocoaPods: GitHub – CocoaPods/Specs: The CocoaPods Master Repo, https://github.com/CocoaPods/Specs/, archived at https://archive.ph/ojsNf
14. Publishing a Crate to Crates.io – The Rust Programming Language, https://doc.rust-lang.org/book/ch14-02-publishing-to-crates-io.html, archived at https://archive.ph/hVUXD
15. De Sutter, B., Schrittwieser, S., Coppens, B., Kochberger, P.: Evaluation Methodologies in Software Protection Research. ACM Computing Surveys **57**(4) (2024). https://doi.org/10.1145/3702314
16. Domínguez-Álvarez, D., de la Cruz, A., Gorla, A., Caballero, J.: LibKit: Detecting Third-Party Libraries in iOS Apps. In: Proc. of the 10th Joint Meeting on Foundations of Software Engineering (ESEC/FSE). ACM (2015). https://doi.org/10.1145/3611643.3616344
17. dontdoit8001: GitHub – [Add] bybit_web3_whitebox_encrypt 1.0.0, https://github.com/CocoaPods/Specs/commit/c9a77987c5739f1c915d4f8b7b41f1c0781a9fc7, archived at https://archive.ph/yOVz2
18. Duan, R., Alrawi, O., Kasturi, R.P., Elder, R., Saltaformaggio, B., Lee, W.: Towards Measuring Supply Chain Attacks on Package Managers for Interpreted Languages. In: Proc. of the 28th Network and Distributed System Security Symposium (NDSS). Internet Society (ISOC) (2021). https://doi.org/10.14722/ndss.2021.23055
19. Fiedler, M.: The Python Package Index Blog – 2FA Requirement for PyPI begins 2024-01-01, https://blog.pypi.org/posts/2023-12-13-2fa-enforcement/, archived at https://archive.ph/Mp08m
20. Fiedler, M.: The Python Package Index Blog – Preventing Domain Resurrection Attacks, https://blog.pypi.org/posts/2025-08-18-preventing-domain-resurrections/, archived at https://archive.ph/JHEX2
21. Fortinet: Supply Chain Attacks: Examples And Countermeasures, https://www.fortinet.com/resources/cyberglossary/supply-chain-attacks/, archived at https://archive.ph/9cJiS
22. GitHub: Configuring two-factor authentication, https://docs.github.com/en/authentication/securing-your-account-with-two-factor-authentication-2fa/configuring-two-factor-authentication, archived at https://archive.ph/vAAUQ
23. GitHub: Upcoming breaking changes for npm v12, https://github.blog/changelog/2026-06-09-upcoming-breaking-changes-for-npm-v12/, archived at https://archive.ph/2DS5s
24. GitHub, Inc: GitHub – [Already shipped] GHCR namespace retirement [GA] (#34085), https://github.com/github/docs/commit/fa650c0a9426f90d68243bde31220c9922b07f50, archived at https://archive.ph/RnFEg
25. GitHub, Inc: GitHub Blog – New tools for open source maintainers, https://github.blog/open-source/maintainers/new-tools-for-open-source-maintainers/, archived at https://archive.ph/AwQcr
26. GoDaddy – Bulk Domain Names – Search and Register Multiple Domains, https://www.godaddy.com/en/domains/bulk-domain-search
27. Google: Google Bug Hunters, https://bughunters.google.com

28. Go Modules Reference – The Go Programming Language, https://go.dev/ref/mod, archived at https://archive.ph/0gYJ2
29. Gradle – Using Resolution Rules, https://docs.gradle.org/current/userguide/resolution_rules.html, archived at https://archive.ph/uV3F3
30. Gu, Y., Ying, L., Pu, Y., Hu, X., Chai, H., Wang, R., Gao, X., Duan, H.: Investigating Package Related Security Threats in Software Registries. In: Proc. of the 44th Symposium on Security & Privacy (S&P). IEEE (2023). https://doi.org/10.1109/SP46215.2023.10179332
31. Guo, W., Xu, Z., Liu, C., Huang, C., Fang, Y., Liu, Y.: An Empirical Study of Malicious Code In PyPI Ecosystem. In: Proc. of the 38th IEEE/ACM International Conference on Automated Software Engineering (ASE) (2023). https://doi.org/10.1109/ASE56229.2023.00135
32. HackerOne, https://hackerone.com
33. Halder, S., Bewong, M., Mahboubi, A., Jiang, Y., Islam, M.R., Islam, M.Z., Ip, R.H., Ahmed, M.E., Ramachandran, G.S., Ali Babar, M.: Malicious Package Detection using Metadata Information. In: Proc. of the World Wide Web Conference (WWW). ACM (2025). https://doi.org/10.1145/3589334.3645543
34. Huang, C., Wang, N., Wang, Z., Sun, S., Li, L., Chen, J., Zhao, Q., Han, J., Yang, Z., Shi, L.: DONAPI: Malicious NPM Packages Detector using Behavior Sequence Knowledge Mapping. In: Proc. of the 33rd USENIX Security Symposium (2024)
35. Huang, Y., Wang, R., Zheng, W., Zhou, Z., Wu, S., Ke, S., Chen, B., Gao, S., Peng, X.: SpiderScan: Practical Detection of Malicious NPM Packages Based on Graph-Based Behavior Modeling and Matching. In: Proc. of the 39th IEEE/ACM International Conference on Automated Software Engineering (ASE). ACM (2024). https://doi.org/10.1145/3691620.3695492
36. Intigriti: Intigriti – Bug Bounty & Agile Pentesting Platform, https://www.intigriti.com
37. Jain, L.: GOPROXY - A Central Module Dependency Management, https://lkumarjain.blogspot.com/2021/08/goproxy-central-module-dependency.html, archived at https://archive.ph/99FEs
38. JitPack – Publish JVM and Android libraries, https://jitpack.io/
39. Kalu, K.G., Singla, T., Okafor, C., Torres-Arias, S., Davis, J.C.: An Industry Interview Study of Software Signing for Supply Chain Security. In: Proc. of the 34rd USENIX Security Symposium (2025)
40. Klivan, S., Höltervennhoff, S., Panskus, R., Marky, K., Fahl, S.: Everyone for Themselves? A Qualitative Study about Individual Security Setups of Open Source Software Contributors. In: Proc. of the 45th Symposium on Security & Privacy (S&P). IEEE (2024). https://doi.org/10.1109/SP54263.2024.00214
41. Laburity Research Team: Research Case Study: Supply Chain Security at Scale – Insights into NPM Account Takeovers, https://labурity.com/research-npm-account-takeovers/, archived at https://archive.ph/R12Lb
42. Ladisa, P., Plate, H., Martinez, M., Barais, O.: SoK: Taxonomy of Attacks on Open-Source Software Supply Chains. In: Proc. of the 44th Symposium on Security & Privacy (S&P). IEEE (2023). https://doi.org/10.1109/SP46215.2023.10179304
43. Ladisa, P., Sahin, M., Ponta, S.E., Rosa, M., Martinez, M., Barais, O.: The Hitchhiker's Guide to Malicious Third-Party Dependencies. In: Proc. of the ACM Workshop on Software Supply Chain Offensive Research and Ecosystem Defenses (SCORED). ACM (2023). https://doi.org/10.1145/3605770.3625212
44. Martínez, J., Durán, J.M.: Software Supply Chain Attacks, a Threat to Global Cybersecurity: SolarWinds' Case Study. International Journal of Safety and Security Engineering **11**(5) (2021). https://doi.org/10.18280/ijsse.110505

45. Maven – Introduction to the Dependency Mechanism, https://maven.apache.org/guides/introduction/introduction-to-dependency-mechanism.html/, archived at https://archive.ph/94Psb
46. maven central repository – Why do I need to verify project ownership?, https://central.sonatype.org/faq/verify-ownership/#answer, archived at https://archive.ph/rS1vp
47. Maven Repository, https://mvnrepository.com/
48. Microsoft: Microsoft Bounty Programs, https://www.microsoft.com/en-us/msrc/bounty
49. Namecheap – Bulk Domain Search – Multiple Domain Checker, https://www.namecheap.com/domains/bulk-domain-search/
50. Neupane, S., Holmes, G., Wyss, E., Davidson, D., Carli, L.D.: Beyond typosquatting: An in-depth look at package confusion. In: Proc. of the 32nd USENIX Security Symposium (2023)
51. Niedziela, G.: YouTube – $130,000+ Learn New Hacking Technique in 2021 – Dependency Confusion – Bug Bounty Reports Explained, https://www.youtube.com/watch?v=zFHJwehpBrU
52. npm – Home, https://www.npmjs.com/, archived at https://archive.ph/ImYpo
53. Ohm, M., Plate, H., Sykosch, A., Meier, M.: Backstabber's Knife Collection: A Review of Open Source Software Supply Chain Attacks. In: Proc. of the 17th International Conference on Detection of Intrusions and Malware, and Vulnerability Assessment (DIMVA). Springer (2020). https://doi.org/10.1007/978-3-030-52683-2_2
54. Ohm, M., Stuke, C.: SoK: Practical Detection of Software Supply Chain Attacks. In: Proc. of the 18th International Conference on Availability, Reliability and Security (ARES). ACM (2023). https://doi.org/10.1145/3600160.3600162
55. Osborne, C.: The Daily Swig – Malicious Python library CTX removed from PyPI repo, https://portswigger.net/daily-swig/malicious-python-library-ctx-removed-from-pypi-repo, archived at https://archive.ph/oLj9G
56. Oversecured: Introducing MavenGate: A Supply Chain Attack Method for Java and Android Applications, https://blog.oversecured.com/Introducing-MavenGate-a-supply-chain-attack-method-for-Java-and-Android-applications/, archived at https://archive.ph/Cj0GE
57. Ozery, Y., Nadler, A., Shabtai, A.: Information Based Heavy Hitters for Real-Time DNS Data Exfiltration Detection. In: Proc. of the 31st Network and Distributed System Security Symposium (NDSS). Internet Society (ISOC) (2024). https://doi.org/10.14722/ndss.2024.24388
58. pip – PyPI, https://pypi.org/project/pip/, archived at https://archive.ph/J7e0L
59. Sammak, R., Rotthaler, A.L., Ramulu, H.S., Wermke, D., Acar, Y.: Developers' Approaches to Software Supply Chain Security: An Interview Study. In: Proc. of the ACM Workshop on Software Supply Chain Offensive Research and Ecosystem Defenses (SCORED). ACM (2024). https://doi.org/10.1145/3689944.3696160
60. Scalco, S., Paramitha, R., Vu, D.L., Massacci, F.: On the feasibility of detecting injections in malicious npm packages. In: Proc. of the 17th International Conference on Availability, Reliability and Security (ARES). ACM (2022). https://doi.org/10.1145/3538969.3543815
61. Schmidt, D., Schrittwieser, S., Weippl, E.: Leaky Apps: Large-scale Analysis of Secrets Distributed in Android and iOS Apps. In: Proc. of the 32nd Conference on Computer and Communications Security (CCS). ACM (2025)

62. Soref, J., wraithgar, drew4237, Karrys, L.: npm Docs – npm install, https://docs.npmjs.com/cli/v11/commands/npm-install, archived at https://archive.ph/5OdJY
63. Statcounter: Mobile Operating System Market Share United States Of America, https://gs.statcounter.com/os-market-share/mobile/united-states-of-america, archived at: https://archive.ph/dfNY1
64. Steinböck, M., Bleier, J., Rainer, M., Urban, T., Utz, C., Lindorfer, M.: Comparing Apples to Androids: Discovery, Retrieval, and Matching of iOS and Android Apps for Cross-Platform Analyses. In: Proc. of the 21st International Conference on Mining Software Repositories (MSR). IEEE/ACM (2024)
65. Swift.org: Swift.org – Package Manager, https://www.swift.org/documentation/package-manager/, archived at https://archive.ph/4i8F5
66. Taylor, M., Vaidya, R., Davidson, D., De Carli, L., Rastogi, V.: SpellBound: Defending Against Package Typosquatting. In: Proc. of the 14th International Conference Network and System Security (NSS) (2020). https://doi.org/10.1007/978-3-030-65745-1_7
67. Therox, O.: CocoaPods Blog – CocoaPods Trunk Read-only Plan, https://blog.cocoapods.org/CocoaPods-Specs-Repo/, archived at https://archive.ph/rs9OM
68. Vu, D.L., Newman, Z., Meyers, J.S.: Bad Snakes: Understanding and Improving Python Package Index Malware Scanning. In: Proc. of the 45th IEEE/ACM International Conference on Software Engineering (ICSE). ACM/IEEE (2023). https://doi.org/10.1109/ICSE48619.2023.00052
69. Williams, J.: What You Need to Know About the SolarWinds Supply-Chain Attack, https://www.sans.org/blog/what-you-need-to-know-about-the-solarwinds-supply-chain-attack/, archived at https://archive.ph/Wb3mn
70. Wyss, E., De Carli, L., Davidson, D.: What the Fork? Finding Hidden Code Clones in npm. In: Proc. of the 44th IEEE/ACM International Conference on Software Engineering (ICSE). ACM/IEEE (2022). https://doi.org/10.1145/3510003.3510168
71. Wyss, E., Wittman, A., Davidson, D., De Carli, L.: Wolf at the Door: Preventing Install-Time Attacks in npm with Latch. In: Proc. of the 17th ASIA Conference on Computer and Communications Security (ASIACCS). ACM (2022). https://doi.org/10.1145/3488932.3523262
72. Zahan, N., Zimmermann, T., Godefroid, P., Murphy, B., Maddila, C., Williams, L.: What are Weak Links in the npm Supply Chain? In: Proc. of the 44th IEEE/ACM International Conference on Software Engineering (ICSE). ACM/IEEE (2022). https://doi.org/10.1145/3510457.3513044
73. Zhang, J., Huang, K., Huang, Y., Chen, B., Wang, R., Wang, C., Peng, X.: Killing Two Birds with One Stone: Malicious Package Detection in NPM and PyPI using a Single Model of Malicious Behavior Sequence. ACM Transactions on Software Engineering and Methodology **34**(4) (Apr 2025). https://doi.org/10.1145/3705304
74. Zhou, X., Zhang, Y., Niu, W., Liu, J., Wang, H., Li, Q.: An Analysis of Malicious Packages in Open-Source Software in the Wild. In: 55th Annual IEEE/IFIP International Conference on Dependable Systems and Networks (DSN) (2025). https://doi.org/10.1109/DSN64029.2025.00045
75. Zimmermann, M., Staicu, C.A., Tenny, C., Pradel, M.: Small World with High Risks: A Study of Security Threats in the npm Ecosystem. In: Proc. of the 28th USENIX Security Symposium (2019)

# A Appendix

## A.1 Detailed Analysis of Other Dependency Systems

### A.1.1 Cargo

*Authentication* The Cargo dependency manager for Rust [6] has a central repository, namely `crates.io`. To publish dependencies, `crates.io` enforces authentication via GitHub. Even if the GitHub namespace hosting the dependency is renamed or deleted, hijacking the published dependency remains impossible because the authentication on `crates.io` is still bound to the original GitHub account. Unlike CocoaPods, `crates.io` stores dependencies directly on its servers rather than forwarding requests to hosting locations [14]. Consequently, to hijack a dependency owner's account, an attacker must hijack the corresponding GitHub account. Since March 2023, GitHub requires 2FA for all users contributing code [22]. Consequently, attackers cannot hijack GitHub accounts, and thus their corresponding `crates.io` accounts, solely through abandoned domains.

A potential attack scenario arises when projects reference dependencies directly via Git URLs. If the referenced URL is hijackable, an attacker can gain control of the repository and inject malicious code. This threat can be partially mitigated by specifying commit hashes or checksums. However, the risk persists during updates, where changes are expected, which can be used to introduce malicious modifications.

*Private Dependencies and Code Execution* Cargo requires developers to specify the dependency registry when using any source other than `crates.io`. This ensures that Cargo fetches the dependency from the correct registry. Furthermore, Cargo does not support arbitrary code execution during dependency installation.

### A.1.2 Go Modules

*Authentication* At first glance, the Go dependency management system appears decentralized because developers specify dependency URLs. However, by default, the origin URL is not fetched directly but rather retrieved via the Go proxy [28,37] that returns the cached library for subsequent invocations. Consequently, the system behaves as a centralized one, and therefore, it is essential to analyze the requirements for publishing new library versions.

The Go proxy does not enforce authentication directly. Instead, it relies on the authentication mechanisms of the platform hosting the library, most commonly GitHub. Gu et al. [30] demonstrated that attackers can hijack dependencies when the dependency hosting URL on GitHub becomes available. In their study, they identified 11,788 dependency URLs available on GitHub. If attackers register an abandoned namespace and image-name, they can publish new, malicious versions. Once the proxy caches these versions, existing projects may unknowingly update to the compromised version. Since developers can reference other third-party URLs, hijacking dependencies also becomes possible once the hosting domain expires. As discussed in Section 2.2, GitHub retires namespace and image-name combinations after they exceed 5,000 downloads to mitigate supply chain attacks. While this mechanism reduces the likelihood of namespace hijacking, it does not fully eliminate the threat, as demonstrated by our analysis of iOS dependencies. Thus, the question arises whether Go dependency caching further undermines this mitigation. Once a dependency is cached, the proxy serves all subsequent downloads rather than GitHub, potentially preventing the dependency from reaching the required download threshold for URL retirement. Additionally, the caching mechanism masks the fact that the original account was deleted or the domain abandoned because the dependency continues to function through the proxy.

In Appendix A.3, we measure the number of dependencies vulnerable to hijacking attacks to better understand how Go's caching mechanism influences the abandonment of GitHub URLs and to quantify dependencies hosted on abandoned domains, both of which were not examined by related work [30].

*Private Dependencies and Code Execution* Developers can configure private proxies for private repositories [28,37]. However, an attacker would still need to hijack the internally used URL to register a private dependency. In addition, Go modules do not support code execution during installation [28]. Consequently, Go is not vulnerable to dependency confusion.

### A.1.3 npm

*Authentication* When a dependency is retrieved from the central npm registry, protection against hijacking attacks relies on npm's authentication mechanisms. In addition to standard username and password authentication, npm recommends enabling 2FA. However, the second factor is delivered via email by default, which does not protect if the associated domain is abandoned. If maintainers do not configure a robust 2FA setup, their dependencies become vulnerable once the associated domain is abandoned [8,41]. Projects can also reference dependencies directly through a URL. In this case, the security responsibility shifts to the platform that hosts the dependency, as with Go and Cargo.

*Private Dependencies and Code Execution* Developers can mix public and private dependencies. Combined with the possibility of executing arbitrary code during dependency installation, it enables dependency confusion attacks [4,30]. Before npm version 12, users could disable script execution during dependency installation through a parameter [62], while the script execution was enabled by default. Starting with version 12, npm disables script execution by default and requires users to enable it explicitly [23], thereby reducing the risk of RCE through dependency confusion.

Similar to the leakage of private CocoaPods dependencies, we observed that iOS app directory structures can expose the npm dependencies used. Unlike typical browser-based applications, where the `node_modules` directory resides on the server and remains hidden from the client, this directory may be included in distributed mobile apps. Consequently, analyzing the app's dependencies can reveal names that may lead to dependency confusion attacks. In Appendix A.2, we analyze the private npm dependencies included in our iOS dataset to gain further insights.

### A.1.4 pip

*Authentication* Hijacking attacks through abandoned email addresses associated with dependency owners have been documented [55]. To mitigate this risk, pip began enforcing 2FA in 2024 for all accounts that publish or modify dependencies [19]. Moreover, starting in 2025, they implemented daily scans for abandoned domains associated with account owners. Once a domain is classified as untrustworthy due to expiration, the system no longer sends password reset emails [20]. This measure reduces the likelihood that an attacker can log in and hijack a dependency after registering an expired domain. Similar to other dependency management systems, pip supports installing dependencies directly from Git repositories or by specifying a URL. In such cases, attackers can again hijack the referenced URLs to inject malicious code.

*Private Dependencies and Code Execution* It is possible to install dependencies from a private repository by specifying the repository via a command-line parameter. However, this configuration remains vulnerable to dependency confusion attacks, as demonstrated by prior work [4].

**A.1.5 Maven** Maven and Gradle are the standard dependency management systems for Java and Kotlin. They typically resolve dependencies through Maven Central [29,45,47], but other public repositories also exist, such as JitPack [38].

*Authentication* Dependencies follow the format `groupId:artifactId:version`. To publish a dependency under a specific `groupId`, e.g., `com.google`, the publisher must prove domain ownership via a DNS record. Dependencies published through source code management platforms such as GitHub also require repository ownership verification [46].

On Maven Central, this verification process is performed automatically or handled by support staff. Consequently, it remains unclear whether previously registered dependencies could be hijacked by acquiring control of an abandoned domain or GitHub namespace, or whether the human verification step prevents such attempts [46,56]. We refrained from testing the hijacking due to ethical concerns. Because multiple public dependency repositories coexist, some projects integrate more than one of them. Attackers can exploit dependencies that do not exist in all repositories with a `groupId` vulnerable to hijacking. They can publish a malicious version of such a dependency to a repository where it previously did not exist. Depending on the resolver configuration, the dependency management system may select and download the malicious artifact instead [30,56].

*Private Dependencies and Code Execution* Similar to Go dependencies, even when private and public repositories are combined, attackers must still gain control of the corresponding hosting URL to publish a malicious dependency in a public dependency repository. Additionally, Maven does not support code execution during installation.

### A.2 npm: Potential Dependency Confusion in Mobile Apps

*Methodology* Similar to our analysis of CocoaPods, we investigated whether dependency names, leaked through the app directory structure, are available in the npm repository.

We extracted all file paths from each app bundle and filtered for paths containing either `node_modules/` or `www/plugins/`. Then, we extracted the directory names immediately following these path segments, as they correspond to npm dependency names. If a directory name started with `@`, indicating an organizational namespace, we also extracted the subdirectory name within that namespace. We excluded all names that contained a dot or an uppercase character, as npm does not allow such characters, leaving these names FPs. We then queried the availability of each extracted name using a `GET` request to `https://registry.npmjs.org/{npm_name}`. This approach can yield FPs for organizational namespaces, since some may be registered to prevent dependency confusion or are used for private dependencies. To eliminate FPs, we manually attempted to register the available namespace. This process does not interfere with name resolution if the corresponding dependency name remains unregistered. Positively, it proactively prevents attackers from claiming the namespace.

We extracted the associated owner email address for existing dependencies to determine whether it pointed to an abandoned domain. We then again used the methodology described in Section 4 to identify abandoned ones.

*Results* Overall, we extracted 671 npm dependency names from 808 iOS apps (8.77%). Among these, 79 dependency names were not registered in the public registry and are vulnerable to dependency confusion attacks if the dependency management file is misconfigured. These vulnerable dependencies appeared in 27 iOS apps, with a median number of installations per app of 500,000 ($MAD =$ 499,500), and one app exceeding 50 million installations. Positively, we did not identify npm dependencies linked to accounts with abandoned email addresses.

### A.3 Effectiveness of GitHub Namespace Retirement for Go

We measure the number of vulnerable open-source projects on GitHub to understand how Go's caching mechanism influences the abandonment of GitHub URLs and whether dependencies are hosted on abandoned domains. Prior work [30] did not analyze these aspects. Understanding the effectiveness of GitHub's mitigation strategies in the Go ecosystem and assessing whether abandoned URLs pose a threat to it are essential to evaluating the robustness of Go's design and developing effective mitigation strategies.

*Methodology* To identify abandoned Go dependencies, we automatically collected all `go.mod` files from public repositories with over 1,000 stars using the GitHub API. With regular expressions, we automatically extracted each dependency's URL and applied our analysis from Section 4 to identify abandoned domains and GitHub namespaces. To eliminate FPs caused by GitHub's namespace retirement mechanisms, we again manually attempted to register each identified namespace and checked whether the image-name is retired, as illustrated in Figure 2. In the second step, we searched all public repositories for usages of vulnerable dependencies via the GitHub API.

*Results* In total, we downloaded 5,848 `go.mod` files from 2,811 repositories belonging to 2,076 GitHub namespaces in March 2025. From these dependency files, we extracted dependencies hosted on 4,520 GitHub namespaces and 219 domains. Among them, 30 namespaces (0.66%) and one domain (0.46%) were available for registration. Our manual verification revealed that seven of the available GitHub namespaces (23.33%), encompassing nine image-names, were vulnerable to hijacking attacks. GitHub's mitigation strategy of retiring URLs effectively prevented hijacking attacks for the remaining ones. Our results show that popular projects are also affected by depending on vulnerable libraries. For instance, `cubefs`, a cloud storage software project with 5,175 stars (as of July 2025), is vulnerable. Projects from well-known companies were also impacted, such as `aws/amazon-ecs-agent` in the case of `deniswernert/udev`, or `kubernetes/ingress-gce` and `kubernetes/test-infra` in the case of the abandoned domain `tmz.dev`.

If attackers hijack any of these libraries, they could compromise the affected projects and devices running them. Our analysis only covers public GitHub projects, yet private projects are also likely to use these vulnerable dependencies. We suspect that the Go module caching mechanism partially undermines GitHub's URL retirement strategy, leaving popular projects vulnerable.

*Mitigation* As an immediate mitigation, we registered the namespaces of all vulnerable image-names and the abandoned domain. This registration does not interfere with any forwarding. Its only purpose is to prevent attackers from performing hijacking attacks. The mitigation only addresses the attack at a single point in time and only for dependencies used by the most popular Go projects. For a sustainable solution, we propose two long-term mitigation strategies: (1) Introduce an ownership mechanism within the Go proxy, similar to Cargo's authentication via GitHub. This mechanism would ensure that only legitimate owners can publish new dependency versions, thereby preventing unauthorized hijacks. (2) Strengthen GitHub's retirement policy by permanently retiring all previously used namespace and image-name combinations, rather than relying on popularity metrics. However, this will not mitigate the issue of abandoned domains and relies on the hosting provider.